\documentclass[aps,pre,reprint]{revtex4-2}
 
\usepackage{blindtext}
\usepackage{amsmath}
\usepackage{amsfonts}
\usepackage{graphicx}
\usepackage{appendix}
\usepackage{color}

\newcommand{\vres}{{v_\mathrm{res}}}
\newcommand{\vthr}{{v_\mathrm{thr}}}
\newcommand{\vmin}{{v_\mathrm{min}}}
\newcommand{\Xr}{{x_\mathrm{r}}}
\newcommand{\Xt}{{x_\mathrm{t}}}
\newcommand{\Xm}{{x_\mathrm{min}}}
\newcommand{\tauARP}{{\tau_0}}

\newcommand{\taum}{{\tau_\mathrm{m}}}

\newcommand{\Cm}{{C_\mathrm{m}}}
\newcommand{\LFP}{\mathcal{L}}
\newcommand{\LFPadj}{\mathcal{L}^\dagger}
\newcommand{\psires}{{\vec{\psi}_\mathrm{res}}}
\newcommand{\lrmean}[1]{{\langle #1 \rangle}}
\newcommand{\Wr}{\mathcal{W}}

\newcommand{\EqRef}[1]{Eq.~\eqref{#1}}

\begin{document}

\title{A `Rosetta stone' for the population dynamics of spiking neuron networks}

\author{Gianni V. Vinci}
\affiliation{Natl. Center for Radioprotection and Computational Physics, Istituto Superiore di Sanità, 00169 Roma, Italy}
\affiliation{PhD Program in Physics, “Tor Vergata” University of Rome, 00133 Roma, Italy}

\author{Maurizio Mattia}
\email{maurizio.mattia@iss.it}
\affiliation{Natl. Center for Radioprotection and Computational Physics, Istituto Superiore di Sanità, 00169 Roma, Italy}

\date{November 2, 2021 -- Ver. 7MM}

\begin{abstract} 
Populations of spiking neuron models have densities of their microscopic variables (e.g., single-cell membrane potentials) whose evolution fully capture the collective dynamics of biological networks, even outside equilibrium.
Despite its general applicability, the Fokker-Planck equation governing such evolution is mainly studied within the borders of the linear response theory, although alternative spectral expansion approaches offer some advantages in the study of the out-of-equilibrium dynamics.
This is mainly due to the difficulty in computing the state-dependent coefficients of the expanded system of differential equations.
Here, we address this issue by deriving analytic expressions for such coefficients by pairing perturbative solutions of the Fokker-Planck approach with their counterparts from the spectral expansion.
A tight relationship emerges between several of these coefficients and the Laplace transform of the inter-spike interval density (i.e., the distribution of first-passage times).
`Coefficients' like the current-to-rate gain function, the eigenvalues of the Fokker-Planck operator and its eigenfunctions at the boundaries are derived without resorting to integral expressions.
For the leaky integrate-and-fire neurons, the coupling terms between stationary and nonstationary modes are also worked out paving the way to accurately characterize the critical points and the relaxation timescales in networks of interacting populations.
\end{abstract}

\maketitle

\section{INTRODUCTION}

As in many other complex systems like macroscopic bodies composed of atoms and molecules \cite{Landau1969}, the collective dynamics of a network of neurons can be effectively described looking at the time evolution of the density function $p$ of its microscopic state variables. 
This density changes in time according to the following continuity equation
\begin{equation}
\partial_t p = -\vec{\nabla} \cdot \vec{S}_p \, ,
\label{eq:Continuity}
\end{equation}
where $\vec{S}_p$ is the net flux of particles, i.e., the probability current of neurons flowing out from an infinitesimal volume of the phase space.
Neurons interact by exchanging action potentials, the so-called `spikes', which are brief and stereotyped messages traveling along their axons.
Given a ionic current $I(t)$ a neuron receives through its membrane, the timing of the spikes it emits can be faithfully reproduced by generalized integrate-and-fire (IF) neuron models \cite{Pozzorini2015,Teeter2018}.
The core state variable of IF neurons is the somatic potential $V(t)$ integrating the input $I(t)$ as a passive RC circuit:
\begin{equation}
\frac{dV}{dt} = F(V) + \frac{I(t)}{\Cm} \, .
\label{eq:VDyn}
\end{equation}
Here $\Cm$ is the membrane capacitance set to 1 by expressing $I(t)$ in units of voltage per time, and $F(V)$ is the drifting current determining the model-specific relaxation dynamics towards the equilibrium potential $V=0$ \cite{Gerstner2014}. 
$F(V)$ incorporates the decay time $\taum$ of the free membrane potential.
When $V(t)$ crosses the threshold value $\vthr$, a spike is emitted and the membrane potential is reset to $V = \vres$ for a refractory period $\tauARP$ past which the subthreshold dynamics \eqref{eq:VDyn} is restored.

Additional state variables such as synaptic conductances or activity-dependent ionic currents can be in principle incorporated to make more realistic the single-neuron spiking activity \cite{Tuckwell1988,Gerstner2014}. 
However, one-dimensional IF neurons preserve without loss of generality, all the key features of the network dynamics of interest here.
In this framework the current $I(t)$ is given by the combination 
$I(t) = \sum_{i=1}^{K} J_i(V) \sum_k \delta(t-t_{i,k} -\delta_i)$, 
where $K$ is the number of presynaptic neurons and $J_i(V)$ is the synaptic efficacy, i.e., the instantaneous change of $V(t)$ each spike emitted at time $t_{i,k}$ by the presynaptic neuron $i$ induces after the axonal delay $\delta_i$.
For mammalian cortical networks it is reasonable to assume a large number of presynaptic contacts ($K \to \infty$) and relatively small excitatory and inhibitory synaptic efficacies ($J_i \to 0$) \cite{DeFelipe2002,Markram2015}.
In this limit, $I(t)$ can be linearized ($\partial_V J_i \simeq 0$) \cite{Amit1991b} and the diffusion limit holds \cite{Tuckwell1988} such that the current is well approximated by a Gaussian white noise with infinitesimal mean $\mu(t)$ and variance $\sigma^2(t)$
\begin{equation}
\begin{array}{rcl}
	     \mu(\nu) & = & K J \nu(t) \\ 
   \sigma^2(\nu) & = & K J^2 (1+\Delta^2) \nu(t)
\end{array} \, .
\label{eq:MuSigma}
\end{equation}
For simplicity's sake, here we considered a homogeneous population of presynaptic neurons with firing rate $\nu(t) = 1/K \sum_{i=1}^K \sum_k \delta(t-t_{i,k} -\delta_i)$ and synaptic efficacies randomly distributed with mean $J$ and standard deviation $\Delta J$.

By incorporating the above stochastic current in \EqRef{eq:VDyn}, a Langevin equation for the membrane potential of a single neuron results:
\begin{equation}
   dV = \left[F(V) + \mu\right]dt + \sigma dW \, .
\label{eq:VDynDiff}
\end{equation}
Here, $W(t)$ is the standard memoryless Wiener process with 0 mean and unit variance: $\langle W(t) W(t')\rangle=\delta(t-t')$.
Note that, as the infinitesimal moments in \EqRef{eq:MuSigma} depend on time via the firing rate $\nu(t)$, $V(t)$ is in general a inhomogeneous stochastic process \cite{Brunel1999,Mattia2002} leading to a faithful description of the out-of-equilibrium population dynamics.
In homogeneous populations of $N$ neurons with similar single-neuron parameters like $\vthr$, $\vres$ and $\taum$, and constant connection likelihood $\epsilon = K/N$, the membrane potentials of different cells can be seen as independent stochastic realizations of the same \EqRef{eq:VDynDiff}.
In the cortical network limit such independence is granted by having $J\to 0$, while the central limit theorem with $K \to \infty$ leads to have input currents with the same $\mu$ and $\sigma^2$.

This `extended' mean-field approximation \cite{Amit1997} including also the fluctuation size $\sigma$ of the `field' (i.e., the synaptic current), is the foundation of the so-called population density approach \cite{Knight1996,Brunel1999,Nykamp2000,Knight2000,Mattia2002} where the continuity equation \eqref{eq:Continuity} reduces to the following Fokker-Planck equation:
\begin{equation}
\partial_{t}p = -\partial_v S_{p}
              = -\partial_{v}[(F+\mu) p] +\frac{1}{2} \sigma^2 \partial^2_v p 
				  \equiv \mathcal{L} \, p \, .
\label{eq:FP}
\end{equation}
Here $\mathcal{L} = -\partial_{v}(F+\mu) +\frac{1}{2} \sigma^2 \partial^2_v$ is the Fokker-Planck operator.
From \EqRef{eq:MuSigma} it is apparent that all neurons are driven by different currents with same statistics determined by the firing rate $\nu(t)$ of the whole population. 
Thus, although two generic neurons are not directly influencing each other, they both have dynamics conditioned by the same collective activity of the network.

In this framework, the firing rate is the flux of realizations (i.e., neurons) crossing the emission threshold $\vthr$ 
\begin{equation}
  \nu(t) = S_p(\vthr) = - \frac{1}{2} \left.\sigma^2 \partial^2_v p\right|_{v=\vthr} \, ,
\label{eq:Nu}
\end{equation}
resulting from \EqRef{eq:FP} by setting $p(\vthr,t) = 0$ as $\vthr$ is an absorbing barrier for the diffusive process $V(t)$ \cite{Tuckwell1988}.

Finding the general solutions of \EqRef{eq:FP} is challenging, and this is due to two main reasons. 
Firstly, the Fokker-Planck operator depends on $\mu(\nu)$ and $\sigma^2(\nu)$, and, as the firing rate depends on $p(v,t)$ from \EqRef{eq:Nu}, it makes $\mathcal{L} = \mathcal{L}(p)$. \EqRef{eq:FP} is thus a nonlinear partial differential equation.
Secondly, when a neuron emits a spike, it does not disappear even if it is `absorbed' in $\vthr$. 
Actually, its dynamics is restored after resetting the membrane potential to $\vres$. 
This guarantees that the number of neurons is conserved, i.e., $\int_\vmin^\vthr{p(v,t) dv} = 1$ at any time $t$.
Such reset is incorporated in \EqRef{eq:FP} by reentering the exiting flux of realizations in $\vthr$ as a point-like source $\delta(v-\vres) \nu(t-\tauARP)$. This source gives rise to a peculiar boundary condition: $S_p(\vres^+)-S_p(\vres^-)=\nu(t-\tauARP)$.
As a result, the flux $S_p(v)$ is non-vanishing even under stationary conditions \cite{Abbott1993,Knight1996,Brunel1999,Fusi1999}, leading to a breaking of the so-called `potential condition' \cite{Risken1984} usually holding in equilibrium statistical physics and hardly relying on the detailed balance. 

In addition to the numerical integration of the partial differential equations \eqref{eq:Continuity} and \eqref{eq:FP} \cite{Nykamp2000,Omurtag2000,deKamps2003,Cai2004}, their analytical characterization is mainly investigated via the linear response theory dealing with the small deviations from the equilibrium states \cite{Abbott1993,Treves1993,Brunel1999,Fusi1999,Knight2000,Brunel2001,Richardson2007,Richardson2010}.
Due to its derivation intricacy only few valuable attempts have been pursued investigating perturbatively nonlinear dynamical regimes like limit cycles \cite{Brunel1999,Brunel2000}.

An alternative approach is to expand the density $p$ as a linear combination of the eigenfunctions $\phi_n(v)$ of the Fokker-Planck operator $\mathcal{L}$
\begin{equation}
\mathcal{L} \phi_n = \lambda_n \phi_n \, ,
\label{eq:EigenSys}
\end{equation}
corresponding to the eigenvalues $\lambda_n$, i.e., the spectrum of $\mathcal{L}$.
Firstly adopted in \cite{Abbott1993,Knight1996} to study the population dynamics of neuronal networks, this is a standard approach in statistical physics \cite{Risken1984,Fox1989}.
It is reminiscent of the time-dependent Hartree-Fock theory in quantum mechanics \cite{McLachlan1964} where the basis $\{\phi_n\}_{n \in \mathbb{Z}}$ moves in time following the system evolution.
Indeed, $\mathcal{L}$ -- and hence $\phi_n$ -- depends parametrically on the time-dependent firing rate \cite{Knight1996,Mattia2002}.
Interestingly, the nonstationary modes $\phi_n$ associated to the non-vanishing eigenvalues $\lambda_n$ of $\mathcal{L}$, display a hierarchy of timescales \cite{Knight1996,Mattia2002,Augustin2017} such that only the slowest modes of the spectrum contribute to the firing rate dynamics making it low-dimensional \cite{Abbott1993,Mattia2002,Schaffer2013}.

This elegant `spectral expansion' approach is particularly well suited to describe the out-of-equilibrium dynamics of neuronal networks as it does not relies on the perturbations of equilibrium states. 
However, it comes at a price: the coefficients of such expansion taking into account recurrent and external synaptic couplings, involve integrals that only in simplified models can be analytically solved \cite{Knight2000,Mattia2002}.
Not only, if the firing rate fluctuates due to a broadband input from upstream neurons -- or endogenously expressed in finite-size networks (i.e.,  composed of a finite number of neurons) -- are taken into account, the approach requires to manage series which are not easily summable.
This is an issue typically arising when second-order statistics like the Fourier power-spectrum $P_\nu(\omega) = |\nu(\omega)|^2$ is studied under stationary conditions \cite{Mattia2002,Mattia2004,Mattia2019}.

Starting from such remarks, here we address the following question: is there any way to exploit the advantages of both the perturbative and the spectral expansion approach to overcome their intrinsic limitations?
In what follows we aim at using the relaxation dynamics of an uncoupled set of neurons and the response to small perturbations of a coupled network as a `Rosetta stone'. 
This will allow us to translate some key expressions with a closed form in both linear response and renewal theory into compact and manageable sum of series and state-dependent coefficients for the spectral expansion approach.
The results of this effort pave the way to a further exploitation of the spectral expansion to investigate the out-of-equilibrium dynamics of spiking neuron networks.

\section{Matching the relaxation of $\nu(t)$}

Under stationary conditions and in presence of synaptic current fluctuations ($\sigma \neq 0$), the collective firing rate $\nu(t)$ of a set of uncoupled IF neurons always approaches an equilibrium point $\nu_0$. 
Due to the reset mechanism following the emission of a spike, IF neurons are Markovian renewal processes. 
As such, inter-spike intervals (ISIs) separating two consecutive spikes do not depend on the ISIs occurred before.
Finding the probability $\rho(t) dt$ to have an ISI in the interval $[t,t+dt]$ is a first-passage time problem \cite{Tuckwell1988}, since it requires to know when $V(t)$ crosses the threshold $\vthr$ for the first time starting from $V(0) = \vres$. 
With this initial condition, the ISI density can be interpreted as the probability of first-spike occurrence $\rho_1(t) \equiv \rho(t)$, leading to recursively define the density $\rho_{k}(t)$ of the $k$-th emitted spike as 
\begin{equation}
	\rho_{k}(t)=\int_{0}^{t}\rho(t)\rho_{k-1}(t-\tau) d\tau \, .
\label{eq:LTofRhoK}
\end{equation}
This `renewal equation' \cite{Tuckwell1988,Gerstner2014} allows to derive the so-called `hazard' function, i.e., the mean density of spikes emitted by an isolated IF neuron at time $t$ 
\begin{equation}
	\nu(t) = \sum_{k=1}^{\infty}\rho_{k}(t) \, ,
\label{eq:NuFromRenewalEq}
\end{equation}
which in turn is equivalent to the relaxation to equilibrium of the firing rate $\nu(t)$.

The firing rate in \EqRef{eq:NuFromRenewalEq} has a straightforward Laplace trasform $\nu(s)\equiv\int_0^\infty \nu(t) e^{-st} dt$. 
Indeed, being $\rho_{k}(t)$ a convolution, \EqRef{eq:LTofRhoK} reduces to
\begin{displaymath}
	\rho_{k}(s) = \rho(s) \rho_{k-1}(s) = \rho(s)^k \, .
\end{displaymath}
Introducing it in \EqRef{eq:NuFromRenewalEq}, $\nu(s)$ results to be
\begin{equation}
\begin{split}
	\nu(s) & = \sum_{k=1}^{\infty}\rho_{k}(s) = \sum_{k=1}^{\infty}\rho(s)^k = \frac{1}{1-\rho(s)} - 1 \\
	                & = \frac{\rho(s)}{1-\rho(s)} \, .
\end{split}
\label{eq:LTofNu}
\end{equation}
This equation establishes a direct nonlocal relationship between the firing rate of a homogeneous pool of independent neurons and the ISI density of a single cell \cite{Cox1977,Gerstner2014}.

\subsection{Relaxation dynamics from the spectral expansion}

The firing rate $\nu(t)$ of the mentioned set of independent neurons can be alternatively derived by expanding the density $p(v,t)$ in \EqRef{eq:FP} as
\begin{displaymath}
	p(v,t)=\sum_n a_n(t) \phi_{n}(v) \, ,
\end{displaymath}
where $a_n(t) = \langle \psi_n|p \rangle = \int_\vmin^\vthr \psi_n(v) p(v,t) dv$ are the projections of $p$ on the eigenfunctions (i.e., the modes) of the operator $\mathcal{L}$ defined in \EqRef{eq:EigenSys}.
The infinite set $\{\psi_n(v)\}$ with $n \in \mathbb{Z}$, is an orthonormal basis of a non-Hermitian space such that $\langle \psi_n|\phi_m \rangle = \delta_{nm}$ \cite{Knight1996,Mattia2002}.
Here, $\psi_n(v)$ are the eigenfunctions of the adjoint operator $\LFPadj = (F+\mu)\partial_{v} +\frac{1}{2} \sigma^2 \partial^2_v$ defined as $\langle \LFPadj \psi_n | \phi_n\rangle \equiv \langle \psi_n | \LFP \phi_n\rangle$, and having the same spectrum as in \EqRef{eq:EigenSys}: $\LFPadj \psi_n = \lambda_n \psi_n$.
The `expanded' $p$ in the Fokker-Planck equation leads to the following equivalent dynamics for the projections $a_n$:
\begin{equation}
\begin{split}
\dot{\vec{a}} & = \mathbf{\Lambda} \vec{a} \\
\nu           & = \Phi + \vec{f}\cdot\vec{a}
\end{split} \, ,
\label{eq:EREuncoupled}
\end{equation}
where the contribution due to the stationary mode ($n = 0$ with $\lambda_0 = 0$) has been isolated, and a matrix formalism has been adopted such that $\{\vec{a}\}_n = a_n$ and $\{\mathbf{\Lambda}\}_{nm}=\lambda_n \delta_{nm}$ with $m,n \neq 0$ \cite{Knight1996,Mattia2002}.
The current-to-rate gain function
\begin{equation}
\Phi(\mu,\sigma) = S_{\phi_0}(\vthr) = -\left.\frac{1}{2}\sigma^2\partial_v^2\phi_0\right|_{v=\vthr}
\label{eq:GainFunc}
\end{equation}
is the flux of realization crossing $\vthr$ under stationary condition, i.e., the asymptotic firing rate $\nu_0$. 
The flux due to the nonstationary modes ($\phi_n$ with $n \neq 0$) are instead the elements of the infinite vector $\vec{f}$, which can be conveniently set to $f_n = 1/\taum$ \cite{Deniz2016}.

The linear system \eqref{eq:EREuncoupled} has a straightforward solution having constant  coefficients ($\mu$ and $\sigma$ do not depend on time):
\begin{equation}
\begin{split}
   \vec{a}(t) & = e^{\mathbf{\Lambda}t} \vec{a}(0) \\
   \nu(t)     & = \Phi + \vec{f} \cdot \vec{a}
\end{split} \, .
\label{eq:NuFromEREuncoupled}
\end{equation}
As all neurons have a starting membrane potential $V(0) = \vres$, $p(v,0) = \delta(v-\vres)$ and the initial value of the projections result to be 
\begin{displaymath}
	a_n(0) = \langle \psi_n|p(v,0)\rangle = \psi_n(\vres) \, .
\end{displaymath}

\subsection{ISI moments and sums in the spectral expansion}

A first sum of coefficients from the spectral expansion results by setting $t=0$ in the relaxation dynamics \eqref{eq:NuFromEREuncoupled}: 
\begin{equation}
   \Phi = - \vec{f} \cdot \psires \equiv - \frac{1}{\taum}\sum_{n\neq 0}{\psi_n(\vres)} \, ,
\label{eq:Nu0asSum}
\end{equation}
where we took into account that $\nu(0) = 0$. 
To our knowledge, this is a novel expression to carry out the current-to-rate gain function $\Phi(\mu,\sigma)$ for any IF neuron model.

We anticipate that this is only a particular case of a more general result.
Indeed, the Laplace transform $\nu(s)$ from the renewal theory in \EqRef{eq:LTofNu} can be directly compared to the one resulting from the firing rate equation \eqref{eq:EREuncoupled}.
In this case we have 
\begin{displaymath}
\begin{split}
   s \, \vec{a}(s) - \vec{a}(0) & = \mathbf{\Lambda} a(s) \\
                     \nu(s)     & = \frac{\Phi}{s} + \vec{f} \cdot \vec{a(s)}
\end{split} \, ,
\end{displaymath}
leading to $\vec{a}(s) = (s \mathbf{I} - \mathbf{\Lambda})^{-1} \psires$, and eventually to
\begin{equation}
   \nu(s) = \frac{\Phi}{s} + \vec{f} \cdot (s \mathbf{I} - \mathbf{\Lambda})^{-1} \psires \, .
\label{eq:LTofNuFromERE}
\end{equation}
Recalling \EqRef{eq:LTofNu} and comparing the two transform $\nu(s)$ we obtain:
\begin{equation}
	\frac{\rho(s)}{1-\rho(s)} - \frac{\Phi}{s} = \vec{f} \cdot (s \mathbf{I} - \mathbf{\Lambda})^{-1} \psires \equiv h(s) \, ,
\label{eq:HofS}
\end{equation}
highlighting a tight relationship between the ISI density $\rho(t)$, the eigenvalues $\lambda_n$ and $\psires$.

From \EqRef{eq:HofS} other sums can be worked out recalling that the ISI moments $\lrmean{(-t)^k} = \lim_{s \to 0}\frac{d^k \rho(s)}{ds^k}$. 
We then perform the same limit on both hand sides of the equation finding that
\begin{displaymath}
   \lim_{s \to 0} h(s) = \frac{\Phi \lrmean{t^2} - 2\lrmean{t}}{2\lrmean{t}} \, .
\end{displaymath} 
Here, considering that the gain function $\Phi = 1/\lrmean{t}$ and that $h(0) = -\vec{f} \cdot \mathbf{\Lambda}^{-1} \psires$, we obtain
\begin{equation}
   h(0) = \frac{c_v^2 - 1}{2} = - \frac{1}{\taum}\sum_{n\neq 0} \frac{\psi_n(\vres)}{\lambda_n} \, ,
\label{eq:Hof0}
\end{equation}
where $c_v = \sqrt{\lrmean{t^2} - \lrmean{t}^2}/\lrmean{t}$ is the coefficient of variation of the ISIs.

We remark that other sums can be worked out deriving by $s$ both hand sides of \EqRef{eq:HofS} and taking the limit $s \to 0$.
Indeed, in this limit the derivatives of $h(s)$ reduce to
\begin{displaymath}
	\left. \frac{d^k h}{ds^k} \right|_{s \to 0}= - k! \vec{f} \cdot \mathbf{\Lambda}^{-(k+1)} \psires = -\frac{k!}{\taum} \sum_{n\neq 0} \frac{\psi_n(\vres)}{\lambda_n^{k+1}}
\end{displaymath}
for any $k > 0$.

\subsection{Eigenvalues $\lambda_k$ and eigenfunction values $\psi_k(\vres)$}

We can further exploit \EqRef{eq:HofS} by making use of the Cauchy's residue theorem.
Indeed, 
\begin{equation}
   \mathrm{Res}_{s=\lambda_k} h(s) = \mathrm{Res}_{s=\lambda_k} \frac{\rho(s)}{1-\rho(s)}=\psi_{k}(\vres)
\label{eq:ResiduePsiK}
\end{equation}
for $\lambda_k \neq 0$.
This allows us to work out from the Laplace transform of the ISI probability density the value of the $k$-th eigenfunction of $\LFPadj$ at $v = \vres$.
In this way $\psi_{k}(\vres)$ can be found without explicitly knowing neither the analytic expression for $\psi(v,s)$ nor the exact value of the $\lambda_k$.
As $s = \lambda_k$ are poles of $h(s)$, a generic line integral of $\rho/(1-\rho)$ around each eigenvalues will give $\psi_{k}(\vres)$.
Similarly, the eigenvalues $\lambda_k$ can be obtained without resorting to a often complicated minimization procedure as
\begin{equation}
   \frac{\mathrm{Res}_{s=\lambda_k} s \, h(s)}{\mathrm{Res}_{s=\lambda_k} h(s)} = 
   \frac{\lambda_k \psi_k(\vres)}{\psi_k(\vres)} = \lambda_k \, .
\label{eq:ResidueLambdaK}
\end{equation}

\subsection{Spectral equation for $\lambda_k$ from $\rho(s)$}

The link between the spectrum of $\LFPadj$ and $\rho(s)$ pointed out by \EqRef{eq:HofS} should not be surprising as $\rho(t)$ is the density of a first-passage time. 
As such, $\rho(t)$ conditioned to have $V(0) = \vres$ and $V(t) = \vthr$, solves the backward Kolmogorov equation \cite{Cox1977,Tuckwell1988}
\begin{displaymath}
	\partial_t \rho = A(\vres) \partial_\vres \rho + \frac{1}{2} B(\vres) \partial_\vres^2 \rho = \LFPadj_\vres \rho \, ,
\end{displaymath}
with boundary condition $\left.\rho(t)\right|_{\vres=\vthr} = \delta(t)$.
In our case of study $A(\vres) = \left.F+\mu\right|_{v=\vres}$ and $B(\vres) = \sigma^2$.
For the sake of clarity, we write explicitly the dependence of the operator $\LFPadj_\vres$ on the initial value of the membrane potential.

Performing the Laplace transform of both hand sides of the above equation we obtain
\begin{displaymath}
	\left.-\rho(t)\right|_{t=0} + s \rho(s) = \LFPadj_\vres \rho(s)
\end{displaymath}
with $\rho(s)$ the Laplace transform of the ISI density implicitly depending on $\vthr$. 
In the case of interest, $\vres < \vthr$ and no spikes are emitted at $t=0$ such that $\left.\rho(t)\right|_{t=0}=0$ and the above equation reduces to
\begin{displaymath}
	(\LFPadj_\vres - s) \rho(\vres,s) = 0 \, ,
\end{displaymath}
where we made explicit the dependence of $\rho$ on the initial potential $\vres$.
As pointed out in \cite{Siegert1951}, this is exactly the same equation for the eigenfunctions $\psi_k(v) = \psi(v,\lambda_k)$ provided that $s = \lambda_k$ leading to the equivalence:
\begin{displaymath}
	\rho(\vres,s) = a(s) \psi(\vres,s) \, .
\end{displaymath}
The arbitrary function $a(s)$ can be derived taking into account the boundary condition $\rho(\vthr,t) = \delta(t)$ leading to $\rho(\vthr,s) = 1$ and hence to $a(s) = 1/\psi(\vthr,s)$, allowing to recover the known relationship \cite{Siegert1951}
\begin{equation}
   \rho(s)=\frac{\psi(\vres,s)}{\psi(\vthr,s)} \, .
\label{eq:RhoFromPsi}
\end{equation}  
Following \cite{Pietras2020}, this equation allows to derive a spectral equation to determine the eigenvalues $\lambda_k$.
Indeed, from the boundary condition associated to the conservation of flux of realizations exiting from $\vthr$ and reentering in $\vres$ after the emission of spike, we have $\psi(\vres,\lambda_k) = \psi(\vthr,\lambda_k)$ \cite{Knight1996,Knight2000,Mattia2002}.
Incorporating this constraint into \EqRef{eq:RhoFromPsi} we eventually have 
\begin{equation}
   \rho(\lambda_k)=1 \, ,
\label{eq:CEfromRho}
\end{equation}  
i.e., yet another way to compute the eigenvalues of the Fokker-Planck operator in \EqRef{eq:EigenSys}.

\subsection{Normalization factor of eigenfunctions $\psi_k(v)$}

As summarized at the beginning of this Section, the spectral expansion of the population density relies on the assumption that the eigenfunctions $\psi_k(v)$ and $\phi_k(v)$ are orthonormal, i.e.,
\begin{displaymath}
	\langle \psi_n|\phi_m \rangle = \int_\vmin^\vthr{\psi_n(v) \phi_m(v) dv} = \delta_{nm} \, .
\end{displaymath}
Such condition is guaranteed by normalizing the generic solution of the equation $(\LFPadj - \lambda_k) \tilde{\psi}_k = 0$ by a factor $Z_k$ in order to have $\psi_k \equiv \tilde{\psi}_k/Z_k$ and
\begin{displaymath}
	Z_k = \langle \tilde{\psi_k}|\phi_k \rangle = \int_\vmin^\vthr{\tilde{\psi}_k(v) \phi_k(v) dv} \, .
\end{displaymath}
This integral expression requires to know explicitly the eigenfunctions of both $\LFP$ and $\LFPadj$. 
Not only, even when they are available, a closed form of this integral is difficult to be obtained.
Hence, its numerical evaluation is required even when the integration domain is infinite, as usually $\vmin \to \infty$, eventually making the computation of $Z_k$ demanding.

Even in this case, a way out to this issue is offered by the residue theorem starting from the knowledge of $\rho(s)$.
Indeed, inserting \EqRef{eq:CEfromRho} into \EqRef{eq:ResiduePsiK} we have 
\begin{displaymath}
   \mathrm{Res}_{s=\lambda_k} \frac{\tilde{\psi}(\vres,s)}{\tilde{\psi}(\vthr,s) - \tilde{\psi}(\vres,s)}= \psi_k(\vres) = \frac{\tilde{\psi}_k(\vres)}{Z_k} \, ,
\end{displaymath}
where we took into account that the normalization factor does not depend on $v$ and thus the equivalence $\psi(\vres,s)/\psi(\vthr,s) = \tilde{\psi}(\vres,s)/\tilde{\psi}(\vthr,s)$ holds. 
Putting out from the residue computation the numerator which becomes $\tilde{\psi}(\vres,s) \to \tilde{\psi}_k(\vres)$, the following expression results
\begin{equation}
\mathrm{Res}_{s=\lambda_k} \frac{1}{\tilde{\psi}(\vthr,s)-\tilde{\psi}(\vres,s)} = \frac{1}{Z_k}
\label{eq:NormFactorPsi}
\end{equation}
giving the normalization factor as a limit expression without both computing any integral and knowing $\phi_k(v)$.

\subsection{ISI density transform for notable neuron models}

The simplest model is the perfect IF (PIF) neuron introduced in \cite{Gerstein1964} in which the membrane potential $V(t)$ is a diffusion Wiener process with drift, i.e., $F(V) = \mu$.
In this model ISI moments are finite only if $\mu > 0$, and the ISI density $\rho(s)$ is an inverse Gaussian whose Laplace transform is \cite{Darling1953,Cox1977,Tuckwell1988}
\begin{equation}
   \rho_\mathrm{PIF}(s) = \mathrm{exp}\left[\frac{\vthr-\vres}{\sigma^2}
	          \left( \mu-\sqrt{\mu^2 + 2 \sigma^2 s} \right)\right] \, .
\label{eq:LTofRhoPIF}
\end{equation}
From this, the first and second moments of the ISI result to be $\lrmean{t} = 1/\Phi = (\vthr - \vres)/\mu$ and $\lrmean{t^2} = \sigma^2/(\mu^2 \Phi)$ \cite{Tuckwell1988}.

A generalization of the PIF neuron model is the one in which a reflecting barrier at $\vmin = 0$ is incorporated, allowing to have finite moments of the ISI also for $\mu \leq 0$.
The model was introduced as a former `neuromorphic' implementation in VLSI circuits of an IF neuron \cite{Mead1989}.
This VLSI IF (VIF) neuron has a closed form of $\rho(s)$ \cite{Fusi1999}:
\begin{equation}
   \rho_\mathrm{VIF}(s) = \frac{\zeta(s) e^\xi}{\zeta(s) \cosh[\zeta(s)] + \xi \sinh[\zeta(s)]}
\label{eq:LTofRhoVIF}
\end{equation}
with $\xi = \vthr \mu/\sigma^2$ and $\zeta(s) = \sqrt{\xi^2 + 2s \vthr/\sigma^2}$, fixing the reset potential at the reflecting barrier $\vres = \vmin = 0$.

The standard, and widely used, leaky integrate-and-fire (LIF) neuron with $F(V) = -V/\taum$ has also a closed form for the Laplace transform  of the ISI density \cite{Siegert1951,Capocelli1971}:
\begin{equation}
   \rho_\mathrm{LIF}(s) = 
	    \frac{\sqrt{e^{\Xr^2}} \mathcal{D}_{-s}(-\sqrt{2} \Xr)}{\sqrt{e^{\Xt^2}} \mathcal{D}_{-s}(-\sqrt{2} \Xt)}
\label{eq:LTofRhoLIF}
\end{equation}
with $\Xt = (\vthr - \mu \taum)/\sigma\sqrt{\taum}$ and $\Xr = (\vres - \mu \taum)/\sigma\sqrt{\taum}$. $\mathcal{D}_{\nu}(z)$ is the parabolic cylinder function solution of the Weber differential equation \cite{OlverNIST2010}.

From \EqRef{eq:CEfromRho}, the eigenvalues for the PIF neuron result to be \cite{Pietras2020}
\begin{equation}
   \lambda_n = -2 \pi^2 \frac{\sigma^2}{(\vthr - \vres)^2} n^2 + i 2\pi \Phi n \, .
\label{eq:LambdaPIF}
\end{equation}
As under strongly drift-dominated regime ($\mu \gg F(\vthr) - \mu$) both LIF and VIF neurons are well approximated by a PIF model, it is not surprising to see that the above expression has been previously worked out in both \cite{Abbott1993,Knight1996,Knight2000} and \cite{Mattia2002}, respectively.
Equation \eqref{eq:LambdaPIF} can then be exploited as an initial guess of $\lambda_n$ in order to identify a suited line integral to work out the actual eigevalues from \EqRef{eq:ResidueLambdaK}.
Note that, in the regimes dominated by the current fluctuations $\sigma$, $\lambda_n$ are no more well approximated by \EqRef{eq:LambdaPIF} as they are real values \cite{Mattia2002,Vinci2021c}.

\section{Linear response as a 'Rosetta stone'}

\subsection{Linear response from spectral expansion}

Under the `extended' mean-field approximation \cite{Amit1997}, the drift and diffusion coefficients of the Fokker-Planck equation for a network of coupled spiking neurons are function of the firing rate $\nu$.
This because both the infinitesimal mean and variance of the synaptic current depend on the network activity: $\mu = \mu(\nu)$ and $\sigma^2 = \sigma^2(\nu)$.
In this general framework the dynamics of the projection coefficients $a_k(t)$ and of $\nu(t)$ result to be \cite{Knight1996,Mattia2002}
\begin{equation}
\begin{split}
\dot{\vec{a}} & = \left(\mathbf{\Lambda} + \mathbf{C}\dot{\nu}\right)\vec{a} + \vec{c} \, \dot{\nu} \\
\nu           & = \Phi + \vec{f} \cdot \vec{a}
\end{split} \, ,
\label{eq:EREcoupled}
\end{equation}
where in addition to the coefficients found in \EqRef{eq:EREuncoupled}, the coupling terms between modes are defined as $\{\mathbf{C}\}_{nm}=\langle \partial_{\nu}\psi_n|\phi_m\rangle$ and $\{\vec{c}\}_n = \langle \partial_{\nu}\psi_n|\phi_0\rangle$ for any $m, n \neq 0$.

The couplings can in general affect both current moments $\mu$ and $\sigma^2$ independently, and perturbations of the input received by the neurons in the network can be a combination of the following changes 
\begin{displaymath}
\begin{split}
   \mu(t) & = \mu_0 + \epsilon(t) \\
   \sigma^2(t) & = \sigma^2_0 + \epsilon(t)
\end{split} \, ,
\end{displaymath}
In what follows, only the input changes $\epsilon(t)$ up to the first order are taken into account (i.e., $o(\epsilon^2) \simeq 0$).

Under this approximation, the Fourier transform of \EqRef{eq:EREcoupled} leads to the Fourier transfer function $H_\epsilon(\omega)$ of the firing rate \cite{Mattia2002,Mattia2004}
\begin{equation}
   H_\epsilon(\omega) \equiv 
	   \frac{\nu(\omega)}{\epsilon(\omega)} = 
		\partial_\epsilon \Phi + i \omega \vec{f} \dot (i\omega\mathbf{I}-\mathbf{\Lambda})^{-1} \vec{c}_\epsilon \, ,
\label{eq:TFofNu}
\end{equation}
where $\nu(\omega) = \int_{-\infty}^\infty{\nu(t) e^{-i\omega t} dt}$ is the Fourier transform of the firing rate fluctuating around $\nu_0 = \Phi(\mu_0,\sigma_0)$, $\epsilon(\omega)$ is the same transform for $\epsilon(t)$ and $\{\vec{c}_\epsilon\}_n =\langle \partial_\epsilon \psi_n|\phi_0 \rangle$. 
With this notation we generalize the derivatives associated to the input changes as $d/d\epsilon$, which eventually should be replaced by either $d/d\mu$ or $d/d\sigma^2$.

\subsection{Eigenfunction $\psi_n$ from Fokker-Planck equation}

The same transfer function $H(\omega)$ can be worked out directly from the Fokker-Planck \EqRef{eq:FP} \cite{Gammaitoni1998}, a perturbative approach adopted in the study of spiking neuron networks (see for instance \cite{Brunel2001,Lindner2001}).
In networks of LIF neurons for which analytical results can be carried out, we start rescaling the membrane potential $V$ as
\begin{equation}
	x = \frac{V -\mu_0}{\sigma_0} \, ,
\label{eq:XfromV}
\end{equation} 
leading to simplify the Fokker-Planck equation \eqref{eq:FP} into:
\begin{displaymath}
   \partial_t p(x,t)= \partial_x(xp)+\frac{1}{2}\partial_x^2 p \equiv \LFP_x p \, .
\label{eq:LFPLIFstandard}
\end{displaymath}
As above, $\mu_0$ and $\sigma_0$ do not depend on time and set the fixed point around which perturbations are studied. 
For the sake of simplicity we set $\taum=1$. 
A different unit time can be adopted simply multiplying by $\taum$ all the variables like $\mu_0$, $\sigma_0^2$, $\nu(t)$ and $\lambda_n$.

In general, the spectral equation $(\LFP_x - \lambda_n)\phi_n=0$ is a second-order differential equation which can be recast into the general form
\begin{equation}
   k_2(x)\frac{d^2\phi_n}{dx^2} + k_1(x)\frac{d\phi_n}{dx} +k_0(\lambda_n) \phi_n =0
\label{eq:2ndOrderDE}
\end{equation} 
where $\phi_n$ result to be a combination of two fundamental (i.e., independent) solutions $f_1(x,\lambda_n)$ and $f_2(x,\lambda_n)$ of the same equation. 
The coefficients $a$, $b$ and $d$ of the combination are determined by solving a linear system determined by the boundary conditions of the problem (see for instance \cite{Deniz2016}):
\begin{displaymath}
   \phi_n(x) = \left\{
   \begin{matrix}
      a \, f_1(x,\lambda) + b \, f_2(x,\lambda) & \Xr < x < \Xt \\ 
      d \, f_2(x,\lambda)                       & x < \Xr
   \end{matrix}
   \right. \, ,
\end{displaymath} 
where we arbitrarily impose $f_2(x,\lambda_n) \to 0$ in the limit $x \to \Xm$. The threshold $x_\mathrm{t}$ and the reset $\Xr$ values comes from \EqRef{eq:XfromV} by setting $V = \vthr$ and $V = \vres$, respectively.

The eigenfunctions $\psi_n(x) = \psi(x,\lambda_n)$ of the adjoint Fokker-Planck operator $\LFPadj_x$ can be derived from the same fundamental solutions of \EqRef{eq:2ndOrderDE}.
They rely also on the Wronskian $\Wr(x)=f_1'(x)f_2(x) -f_1(x)f'_2(x)$ which can be in general expressed as \cite{OlverNIST2010}
\begin{equation}
	\Wr(x) = C \, e^{-\displaystyle\int \frac{k_1(x)}{k_2(x)}dx} \, ,
\label{eq:Wronskian}
\end{equation}
being a solution of the differential equation $\Wr'(x) = -k_1(x)/k_2(x)$ solvable up to a constant factor $C$.
Note that although $\Wr$ depends on $f_1$ and $f_2$ it does not depend on $\lambda_n$ and for the LIF neuron we have $\Wr_\mathrm{LIF}(x)=e^{-x^2}$. 

As mentioned above, to derive the adjoint eigenfunctions two boundary conditions must be taken into account: i) $\psi_n(\Xr) = \psi_n(\Xr)$ and ii) $\psi_n(x)$ must be smooth on all the domain of $x$ \cite{Knight1996,Knight2000,Mattia2002}.
With these conditions the adjoint eigenfunctions are in general
\begin{equation}
   \psi_n(x) = \psi(x,\lambda_n) = \frac{f_2(x,\lambda_n)}{Z_n \, \Wr(x)} \, ,
	\label{eq:Psi}
\end{equation}
being it a solution of the spectral equation $(\LFPadj_x - \lambda_n)\psi_n = 0$ if \EqRef{eq:Wronskian} is also taken into account \cite{Deniz2016}.
The normalization factor $Z_n$ is the same as the \EqRef{eq:NormFactorPsi}.

\subsection{Linear response from Fokker-Planck equation}

To derive a closed formula for the transfer function $H_\epsilon(\omega)$ directly from the Fokker-Planck equation we follow the perturbative approach exploited in \cite{Brunel1999,Richardson2007} to the study of IF neuron networks. 
In this framework \EqRef{eq:FP} with the standardized variable $x$ can be decomposed by developing in Taylor series the probability density 
\begin{displaymath}
   p(x,t) = \phi_0(x) + \epsilon(t) p_1(x) + o(\epsilon^2) \, .
\end{displaymath} 
The eigenfunction $\phi_0$ is the stationary probability density when the input current has moment $\mu_0$ and $\sigma_0$.
As a consequence, the firing rate $\nu(t)$ is
\begin{displaymath}
	\nu(t) = \Phi + \frac{1}{2\sigma_0} p_1'(x_\mathrm{t}) \epsilon(t) + o(\epsilon^2) \, ,
\end{displaymath}
from which the transfer function we are looking for results to be
\begin{equation}
   H_\epsilon^{(P)}(\omega) = 
      \frac{\nu(\omega)}{\epsilon(\omega)} = 
      \frac{p_1'(x_\mathrm{t})}{2\sigma_0} \, .
\label{eq:TFofNuFromFPE}
\end{equation}
Note that the factor $1/\sigma_0$ comes from having adopted $x$ as state variable, and due to this $\partial_v = 1/\sigma_0 \partial_x$.

Considering that also the operator $\LFP_x$ depends on $\epsilon$ we can write it as $\LFP_x = \LFP_{x0} + \epsilon(t)\LFP_{x1} + o(\epsilon^2)$ leading to
\begin{displaymath}
\begin{split}
	\dot{\epsilon} p_1 & = \LFP_{x0} \phi_0 + (\LFP_{x0} p_1 + \LFP_{x1} \phi_0) \epsilon + o(\epsilon^2) \\
	& \simeq (\LFP_{x0} p_1 + \LFP_{x1} \phi_0) \epsilon
\end{split} \, .
\end{displaymath}
The Laplace transform of both hand side of this equation affects only $\epsilon(t)$, leading to the disappearance of its transform $\epsilon(s)$:
\begin{equation}
	s \, p_1(x) = \LFP_{x0} p_1(x) + F_0(x)
\label{eq:LFP1}
\end{equation}
with $F_0(x) = \LFP_{x1} \phi_0$. 
As in \cite{Brunel1999}, from this equation $p_1(x)$ can be derived as the sum of two solutions: the homogeneous equation $(\LFP_{x0} - s) p_1 = 0$, which is the same as our usual spectral equation with $s = \lambda_n$, and a particular solution $Q(x,s)$ to be determined:
\begin{displaymath}
   p_1(x) = 
      \left\{\begin{matrix}
         a \, f_1(x) + b \, f_2(x) + Q(x)   &   \Xr < x < \Xt \\ 
         d \, f_2(x) + Q(x)                 &   x < \Xr
      \end{matrix}\right. \, .
\end{displaymath}
For the sake of clarity, the dependency on $s$ here is implicit.
As in the above derivation, the usual boundary conditions lead to a system of linear equations in $a$, $b$ and $d$ to be solved. 
After some algebra and referring to \EqRef{eq:Psi} (see Appendix~\ref{sec:AppendixA} for details), we found that:  
\begin{equation}
   p_1'(\Xt) = 
      \frac{\displaystyle\frac{\Wr_{Q,f_2}(\Xt)}{\Wr(\Xt) Z}-\frac{\Wr_{\Delta Q,f_2}(\Xr)}{\Wr(\Xr) Z}}{\psi(\Xt)-\psi(\Xr)}
\label{eq:dP1}
\end{equation}
where the Wronskian $\Wr(x)$ is from \EqRef{eq:Wronskian}, $\Delta Q(x) = \lim_{\delta \to 0}Q(x+|\delta|)-Q(x-|\delta|)$ and $\Wr_{f,g}(x) = f'(x) g(x) - f(x) g'(x)$.
We remark that this formula is valid in general for any IF neuron model.
Besides it depends only on the particular solution $Q(x)$ and on the eigenfunction $\psi$, as from \EqRef{eq:Psi} $f_2(x) = \psi(x) \, Z \, \Wr(x)$ with $Z(s) = Z_n|_{\lambda_n = s}$.
As above all these function implicitly depend also on $s$ and then on the eigenvalues if $s = \lambda_n$.

\subsection{Matching the transfer functions to derive $c_{n0}$}

Focusing now on the specific case of the LIF neuron and considering an input change affecting only the drift $\mu$, we have $F_0(x)= -\phi_0'(x)/\sigma_0$. 
The related particular solution of \EqRef{eq:LFP1} can be verified to be \cite{Brunel1999}
\begin{displaymath}
   Q_{\mu}(x,s) = -\frac{\phi_0'(x)}{\sigma_0(s + 1)} \, .
\end{displaymath}
Recalling that $\nu_0 = \Phi = -\phi_0'(\Xt)/2$, we remark that $Q_{\mu}(\Xt,s)=2\nu_0/[\sigma_0(s + 1)]$ highlighting the direct link between $Q(x)$ and the flux of realizations $S_{\phi_0}(x)$ under stationary condition.
Making use of all this information in \EqRef{eq:dP1} (see Appendix~\ref{sec:AppendixB} for details), we can derive $p_1'$ and hence from \EqRef{eq:TFofNu}, the transfer function of the modulated drift $\mu(t)$ of the input current
\begin{equation}
   H_\mu^{(P)}(s) = \frac{\nu_0}{\sigma_0(s+1)}\frac{\partial_x \psi(\Xt,s)-\partial_x \psi(\Xr,s)}{\psi(\Xt,s)-\psi_(\Xr,s)} \, .
\label{eq:Hmu}
\end{equation}
With a similar derivation detailed in Appendix~\ref{sec:AppendixB}, the transfer function of the perturbations to the variance $\sigma^2(t)$ result to be
\begin{equation}
   H_\sigma^{(P)}(s)=
\frac{\nu_0}{2\sigma_0^2 (s+2)}\left(2- \frac{\Xt\partial_x \psi(\Xt,s)-\Xr\partial_x \psi(\Xr,s)}{\psi(\Xr,s)-\psi_(\Xr,s)}\right)
\label{eq:Hsigma}
\end{equation}
As under the mean-field approximation we know that both the mean and the variance of the input current depend on the firing rate $\nu$, from these expressions the rate-to-rate transfer function $H_\nu(s)$ can be eventually carried out:
\begin{displaymath}
	H_\nu(s) = H_\mu(s) \frac{d\mu}{d\nu} + H_\sigma(s) \frac{d\sigma^2}{d\nu}
\end{displaymath}

We conclude deriving an expression for the coupling coefficients between nonstationary and stationary modes $c_{n0}^{(\epsilon)} = \{\vec{c}_\epsilon\}_n$ found in Eqs.~\eqref{eq:EREcoupled} and \eqref{eq:TFofNu}.
To this purpose we note that the transfer function $H_\epsilon(s)$ (where $s = i\omega$) derived with the spectral expansion approach, can give
\begin{displaymath}
	\mathrm{Res}_{s=\lambda_k} H_\epsilon(s) = \lambda_n \, c_{n0}^{(\epsilon)} \, .
\end{displaymath}
Due to the equivalence between this $H_\epsilon(s)$ and $H_\epsilon^{(P)}(s)$ from the above perturbative approach, we can carry out an explicit expression for the coupling coefficients as
\begin{displaymath}
	c_{n0}^{(\epsilon)} = \frac{1}{\lambda_n} \mathrm{Res}_{s=\lambda_k} H_\epsilon^{(P)}(s) \, .
\end{displaymath}
To compute the residues from Eqs.~\eqref{eq:Hmu} and \eqref{eq:Hsigma}, we point out that the spectral equation \EqRef{eq:CEfromRho} can be expressed directly in terms of adjoint eigenfunction: $\psi(\vres,\lambda_k) = \psi(\vthr,\lambda_k)$.
From this we have
\begin{displaymath}
	\mathrm{Res}_{s=\lambda_k} \frac{1}{\psi(\vthr,\lambda_k) - \psi(\vres,\lambda_k)} = 1 \, ,
\end{displaymath}
which straightforwardly leads to 
\begin{equation}
   c^{(\mu)}_{n0} =
	   \frac{\nu_0}{\sigma_0 \lambda_n(\lambda_n+1)}\left(\psi_n'(\Xt) - \psi_n'(\Xr)\right)
	\label{eq:Cn0Mu}
\end{equation}
and to
\begin{equation}
c^{(\sigma)}_{n0} = 
   - \frac{\nu_0}{2\sigma_0^2 \lambda_n(\lambda_n+2)}\left(\Xt \psi_n'(\Xt) - \Xr \psi_n'(\Xr)\right) \, .
   \label{eq:Cn0Sigma}
\end{equation}
Remarkably, as for the LIF neuron we have
\begin{equation}
   \psi_n(x) = \frac{e^{x^2/2}}{Z_n} \mathcal{D}_{-\lambda_n}(-\sqrt{2} x) \, ,
\label{eq:PsiLIF}
\end{equation}
once we known the eigenvalues $\lambda_n$, the expressions for both $c^{(\mu)}_{n0}$ and $c^{(\sigma)}_{n0}$ are closed formulas, not requiring any integral on the $x$-domain.
Finally, recalling the dependence on $\nu$ of the moments of synaptic current in \EqRef{eq:MuSigma}, the coupling coefficients in the firing rate equation \EqRef{eq:EREcoupled} are
\begin{equation}
\begin{split}
   c_{n0} & = \langle\partial_\nu\psi_n|\phi_0\rangle \\
	       & = \frac{d\mu}{d\nu} c^{(\mu)}_{n0} + \frac{d\sigma^2}{d\nu} c^{(\sigma)}_{n0} \\
			 & = \taum K J \left(c^{(\mu)}_{n0} + J(1+\Delta^2)c^{(\sigma)}_{n0}\right)
\end{split} \, ,
\label{eq:Cn0}
\end{equation}
where we reintroduced explicitly $\taum$.

\section{Conclusion}

Here, by carrying out and pairing the dynamics of the firing rate $\nu(t)$ 
derived from two different approaches, we found novel analytic expressions for 
the coefficients underlying the spectral expansion of the population density $p(v,t)$. 
In the case of an uncoupled set of spiking neurons, we used the renewal theory in combination with the spectral expansion approach by pairing the relaxation dynamics of $\nu(t)$ -- part of our `Rosetta stone' -- eventually obtaining sums of series in closed form. 
More specifically, we uncovered a new expression of the current-to-rate gain function $\Phi(\mu,\sigma)$ in \EqRef{eq:Nu0asSum} as series of the adjoint eigenfunctions evaluated at the reset potential $\psi_n(\vres)$. 
We also found a tight relationship between the moments of the ISI distribution and a suited combination of the $\psi_n(\vres)$ and the eigenvalues $\lambda_n$ of the Fokker-Planck operator. 
We remark that all these findings are model-independent highlighting the descriptive power of the spectral expansion and further extending recent results reported in \cite{Pietras2020}. 

From the same side of our Rosetta stone, we found an alternative way to compute both $\psi_n(\vres)$ and $\lambda_n$. 
Indeed, relying on the residue theorem, Eqs. \eqref{eq:ResiduePsiK} and \eqref{eq:ResidueLambdaK} show how to obtain from the Laplace transform $\rho(s)$ of the ISI density both these coefficients.
Notably, these equations provide a great advantage in practical terms to estimate numerically the eigenvalues $\lambda_n$ even when their analytic expressions are not available.
This is due to the fact that usually eigenvalues results from the numerical search of the roots of the spectral equation \eqref{eq:CEfromRho} \cite{Mattia2002,Pietras2020}, or of the zeros of numerically derived functions \cite{Ostojic2011a,Augustin2017}.
Evaluating residues instead requires computation of a limit or of a line integral around the border of a domain centered around some rough guess for $\lambda_n$.
The expected computational advantages resulting from this residue-based approach can pave the way to effective implementations of the spectral expansion formalism in the numerical integration of the network dynamics of spiking neurons.

For the specific case of LIF neurons, we also derived additional analytic expressions for the coupling coefficients $c_{n0}$ involved in the evaluation of both the out-of-equilibrium dynamics and the critical points of coupled networks via the firing rate equation \eqref{eq:EREcoupled}.
Here we resorted to the linear response to a small perturbation of the input as the other side of our Rosetta stone.
Pairing the perturbative response carried out from the spectral expansion approach and directly from the Fokker-Planck equation, we finally obtained Eqs.~\eqref{eq:Cn0Mu} and \eqref{eq:Cn0Sigma}.
These expressions for $c_{n0}$ result to involve only the eigenfunctions at the reset potential $\psi_n(\vres)$ and the eigenvalues $\lambda_n$.
These coefficients in turn appear to be tightly linked in \EqRef{eq:HofS} to the ISI density $\rho(t)$ of isolated cells.
In other words the coupling coefficients are the expression of a single-neuron feature valid under stationary condition, rather than being a direct function of the synaptic efficacy $J$.
Although this may appear as a contradiction, we remark that in \EqRef{eq:Cn0} these single-neuron features in $c_{n0}^{(\epsilon)}$ mainly play a modulatory role of the synaptic coupling $KJ$, which is fully taken into account in the mean-field expressions for the current moments $\mu(\nu)$ and $\sigma^2(\nu)$.

In the framework of the spectral expansion of $p(v,t)$, approximated expressions for $c_{n0}^{(\mu)}$ has been previously derived in \cite{Knight1996}, finding
\begin{displaymath}
	c_{n0}(\mu) \simeq \frac{\nu_0}{\lambda_n + 1} C(\mu) 
\end{displaymath}
for networks of LIF neurons working in drift-dominated regime ($\mu \taum > \vthr$) and with small synaptic noise ($\Xt \gg 1$).
This is perfectly compatible with what we derived here and reported in \EqRef{eq:Cn0Mu}, provided that $C(\mu) = \left(\psi_n'(\Xt) - \psi_n'(\Xr)\right)/(\sigma_0 \lambda_n)$.
Note that our result has a more general applicability being valid also under subthreshold noise-dominated regime ($\mu \taum < \vthr$).
Another expression for $c_{n0}^{(\mu)}$ has been recently presented also in \cite{Pietras2020}.
In this case the spectral expansion approach targeted the refractory density, leading to obtain for a network of renewal neurons with Poissonian ISIs and a refractory period $\tau_0$, the following expression
\begin{displaymath}
	c_{n0}(\mu) = \frac{\partial_\mu \Phi}{\lambda_n + (1-\tau_0\lambda_n) \nu_0}
	            \xrightarrow[\tau_0 \to 0]{} \frac{\partial_\mu \Phi}{\lambda_n + \nu_0} \, .
\end{displaymath}
This equation appears to be qualitatively different from our \EqRef{eq:Cn0Mu}.
A possible explanation of such disagreement on one hand may reside in the fact that LIF neurons are not Poissonian processes.
On the other hand, this might highlight a qualitative difference in the population dynamics described by the refractory density approach compared to the classical one focused on the evolution of the membrane potential density.

Note that, our derivation of the coupling terms $c_{n0}$ for LIF neurons can in principle be applied to other IF neuron models.
We expect in this case qualitatively similar results, confirming that the ISI density-based coefficients $c_{n0}^{(\epsilon)}$ eventually modulate the system response with strength proportional to $KJ$. 
We also draw the reader attention to the fact that here we did not report any expression for the coupling coefficients $C_{nm}$ between nonstationary modes ($n,m \neq 0$).
They are involved in the nonlinear response of the neuronal network playing a role when the time derivative $\dot{\nu}$ of the firing rate is relatively large and the density $p(v,t)$ is significantly different from the stationary one ($\phi_0$) \cite{Mattia2002}.
Provided that such constraints are not violated, the network dynamics can be fully described by taking into account the only coefficients of the spectral expansion we studied, even when the firing rate is driven outside equilibrium \cite{Augustin2017,Mattia2021b}.   

Regarding the applicability to the out-of-equilibrium dynamics, we further point out that all the mentioned coefficients depend on the infinitesimal moments $\mu$ and $\sigma$ of the synaptic current. 
This means that under the extended mean-field approximation that leads to \EqRef{eq:MuSigma}, all the coefficients depend from time to time on the activity $\nu(t)$ of the network. 
Here $\mu$ and $\sigma$ must be interpreted as parameters. 
Indeed, it is true that our Rosetta stone approach allowed us to carry out these spectral expansion coefficients from the quasi-equilibrium dynamics (linear response and relaxation in uncoupled sets), i.e., considering $\mu$ and $\sigma$ as fixed.
Nonetheless, for a given $\mu$ and $\sigma$ they contribute to faithful represent a snapshot of the moving basis $\left\{|\phi_n\rangle\right\}_{n \in \mathbb{Z}}$ onto which the density $p(v,t)$ is projected.
The basis moves as the drift and diffusion terms in the Fokker-Planck equation \eqref{eq:FP} vary describing a sequence of stochastic processes (i.e., the membrane potentials of the neurons) locally homogeneous in time. 
We believe this is an important point to stress. 
Starting from the linearizable (quasi-equilibrium) dynamics of spiking neuron networks, the spectral expansion coefficients can be derived paving the way to describe the same system outside equilibrium.

\appendix

\section{Derivation of \EqRef{eq:dP1}}
\label{sec:AppendixA}

To derive the formula for the perturbation of the density $p_1(\Xt)$ we need first to find the following coefficients $a$, $b$ and $d$:
\begin{displaymath}
   p_1(x) = 
      \left\{\begin{matrix}
         a \, f_1(x) + b \, f_2(x) + Q(x)   &   \Xr < x < \Xt \\ 
         d \, f_2(x) + Q(x)                 &   x < \Xr
      \end{matrix}\right. \, .
\end{displaymath}
They can be evaluated by taking into account the boundary conditions of the Fokker-Planck equation:
\begin{displaymath}
\begin{split}
a f_1(\Xt) + b f_2(\Xt) & = -Q(\Xt)  \\
a \left[f'_1(\Xt)-f'_1(\Xr)\right] + & \\
b \left[f'_2(\Xt)-f'_2(\Xr)\right] + d \, f'_2(\Xr) & = \Delta Q'(\Xr)-Q'(\Xt)  \\
a \, f_1(\Xr) + b \, f_2(\Xr) - d \, f_2(\Xr) & = -\Delta Q(\Xr)
\end{split} \, .
\end{displaymath}
We note that the homogeneous system is equivalent to the eigenfunction problem. 
Indeed, the setting of the associated determinant to zero gives back the spectral equation, i.e. the boundary condition $\psi_n(\Xr) = \psi_n(\Xr)$, for the eigenvalues $\lambda_n$ \cite{Deniz2016}. 
Solving the above system we have
\begin{widetext}
\begin{displaymath}
a = \frac{Q(\Xt) f_2(\Xr) f'_2(\Xt) - f_2(\Xt) 
       \left[\Delta Q(\Xr) f'_2(\Xr) + f_2(\Xr) 
		   \left(Q'(\Xt) -\Delta Q'(\Xt)
			\right)
       \right]}{\Wr(\Xt) f_2(\Xr) -\Wr(\Xr) f_2(\Xt)}
\end{displaymath}
and
\begin{displaymath}
b = -\frac{f_2(\Xr) 
       \left[Q_(\Xr) \left(f'_1(\Xr) - f'_1(\Xt)\right) + 
             f_1(\Xt) \left(Q'(\Xt) - \Delta Q'(\Xr)\right)\right] + 
       f'_2(\Xt) \left[\Delta Q(\Xr) f_1(\Xt) - Q(\Xt) f_1(\Xr)\right]}
		{\Wr(\Xt) f_2(\Xr) - \Wr(\Xr) f_2(\Xt)} \, .
\end{displaymath}
\end{widetext}
These coefficients determine $p_1(x)$ in the domain of interest ($x > \Xr$), this is why we do not report the expression for $d$.
Evaluating the derivative $p_1'(\Xt)$ of the above density perturbation at the threshold potential eventually leads to \EqRef{eq:dP1}.

\section{Derivation of the transfer function $H^{(P)}_\epsilon$}
\label{sec:AppendixB}

From \EqRef{eq:TFofNuFromFPE} we know that the generic transfer function $H^{(P)}_\epsilon$ is proportional to $p_1'(\Xt)$ given by \EqRef{eq:dP1}, whose derivation has been detailed in the previous Appendix.
Starting from this, here we provide the key steps needed to obtain the explicit expressions for $H^{(P)}_\epsilon$ reported in Eqs.~\eqref{eq:Hmu} and \eqref{eq:Hsigma} for the specific case of the standard LIF neuron model.
If the input modulation involves only the mean $\mu$, $Q_\mu(x) \propto \phi'_0(x)$.
As in the main text, here we assume an implicit dependence on $s$. 
Remembering that $\LFP_{x0} \phi_0 = \partial_x (x \phi_0) + \frac{1}{2}\partial_x^2\phi_0 = 0$ we can write 
\begin{displaymath}
   Q'_\mu(x) = - 2 x \,Q_\mu(x) - 2 \phi_0(x) = \frac{\Wr'(x)}{\Wr(x)}\,Q_\mu(x) - 2 \phi_0(x) \, ,
\end{displaymath}
where we taken into account that from \EqRef{eq:Wronskian} the Wronskian for the LIF neuron reduces to the differential equation $\Wr'(x) =-2x \, \Wr(x)$. 
The first term of the numerator in \EqRef{eq:dP1} then reduces to
\begin{displaymath}
\begin{split}
	\frac{\Wr_{Q,f_2}(\Xt)}{\Wr(\Xt) Z} 
      & = \left.\frac{Q'_\mu f_2 - Q_\mu f_2}{\Wr Z}\right|_{x=\Xt} \\
      & = \left.-\frac{Q_\mu}{Z} \left(- \frac{\Wr' f_2}{\Wr^2} +\frac{f'_2}{\Wr} \right) -2\phi_0 \frac{f_2}{\Wr Z}\right|_{x=\Xt} \\
      & =\left.-\left(Q_\mu +2\phi_0\right) \psi'\right|_{x=\Xt} \\
      & = -Q_\mu(\Xt) \psi'(\Xt)
\end{split}
\end{displaymath}
where we used \EqRef{eq:Psi} replacing $\lambda_n$ with $s$ and defining $Z(s) = Z_n|_{\lambda_n = s}$, and we took into account the boundary condition $\phi_0(\Xt)=0$.
The same can be done with the other term $\Wr_{\Delta Q,f_2}(\Xr)/\Wr(\Xr) Z$ by taking into account the other boundary condition about the flux conservation, $Q(x_t)=\Delta Q(x_r)$, eventually leading to rewrite \EqRef{eq:dP1} as
\begin{displaymath}
p'_1(\Xt) = -Q_\mu(x_t)\frac{\psi'(x_t)-\psi'(x_r)}{\psi(x_t)-\psi(x_r)} \, .
\end{displaymath}
Analogous results hold whenever the particular solution is proportional to the derivative of the stationary eigenfunction. 

In the case of a changing variance $\sigma^2$ of the input current, the forcing term is $F_{0}(t)=\frac{\nu_1(t)}{\sigma_0^2} \phi''_0 $. 
The particular solution is then
\begin{displaymath}
   Q^{\sigma}(x) = K(s) \phi''_0(x) = \frac{\gamma(s)}{2(s+2)}\phi''_0(x) \, .
\end{displaymath}
Using the stationary equation $\LFP_{x0} \phi_0 = 0$, we have that 
\begin{displaymath}
   \partial^3_x \phi_0 = \left(-2 x \partial^2_x-4\partial_x\right) \phi_0 = (4x^2-4)\phi'_0 \, .
\end{displaymath}
By making use of these expressions in \EqRef{eq:dP1}, we eventually obtain for the first term of the numerator 
\begin{displaymath}
   \frac{\Wr_{Q,f_2}(x)}{\Wr(x) Z} = \frac{K \phi'_0}{\Wr Z}\left( -4 f_2 +2x(2x f_2 + f'_2) \right) \, .
\end{displaymath}
As above, the differential definition of $\Wr$ and \EqRef{eq:Psi} lead us to derive the results presented in the main text.
Note that, in order to obtain the transfer function $H^{(P)}_\epsilon$ the complete definition of the variation of the firing rate is $\nu_1 =-\frac{1}{2}p_1'(x)$.


\end{document}